# Plasmonic nano-optical trap stiffness measurements and design optimization

Quanbo Jiang,[1] Jean-Benoît Claude,[1] and Jérôme Wenger[1,*]

[1]*Aix Marseille Univ, CNRS, Centrale Marseille, Institut Fresnel, 13013 Marseille, France*

*jerome.wenger@fresnel.fr

**Abstract:** Plasmonic nano-optical tweezers enable the non-invasive manipulation of nano-objects under low illumination intensities, and have become a powerful tool for nanotechnology and biophysics. However, measuring the trap stiffness of nanotweezers remains a complicated task, which hinders the development of plasmonic trapping. Here, we describe an experimental method to measure the trap stiffness based on the temporal correlation of the fluorescence from the trapped object. The method is applied to characterize the trap stiffness in different double nanohole apertures and explore the influence of their design parameters in relationship with numerical simulations. Optimizing the double nanohole design achieves a trap stiffness 10× larger than the previous state-of-the-art. The experimental method and the design guidelines discussed here offer a simple and efficient way to improve the performance of nano-optical tweezers.

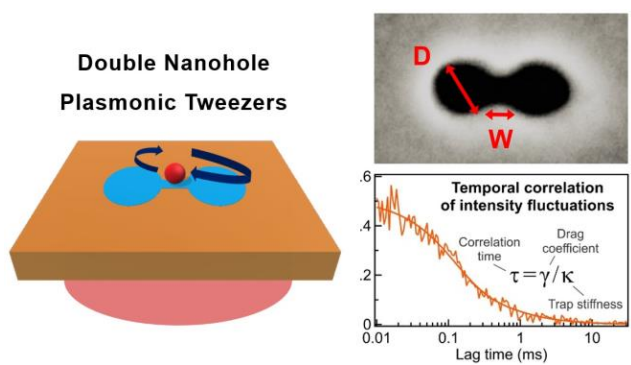

Figure for Table of Contents

## Introduction

Plasmonic nanostructures can generate intense electromagnetic field gradients over subwavelength dimensions, [1] allowing to trap nano-objects that would otherwise be too small or too transparent to be manipulated using conventional diffraction-limited optical tweezers [2–6]. Plasmonic nano-optical tweezers have emerged as a key enabling technology to trap nanoparticles [7–13], quantum dots [14,15], and proteins [16–19] using various plasmonic designs such as single apertures [3], double nanohole apertures [5,20,21], bowtie structures [22–24], coaxial apertures [25,26] or dimer block antennas [27,28].

A central question in plasmonic nano-tweezers goes with the quantification of the trap stiffness. Owing to the nanoscale dimensions of the trap, measuring the stiffness is not an easy task. One strategy implies measuring the transmission signal in the presence of the nanoparticle and computing its temporal correlation [7]. A closely related approach computes the power spectral density of the transmitted signal [22]. Alternatively, the evolution of the transmission at the beginning of the trapping event also contains information about the trap stiffness [7,11]. One drawback associated with the measurement of the transmission is that it works over a bright background level, so that the occurrence of a trapping event is only associated with a minor change of the transmission by a few percent. Moreover, this approach features a reduced contrast near the plasmonic resonance of the structure [22]. For plasmonic nanoantennas where the position of the trapped nanoparticle can be imaged on a camera, the trap stiffness can be deduced from the histogram of the nanoparticle positions [4,15,27,29]. However, this imaging approach is challenging to implement as it requires nanoscale localization accuracy and bright photon fluxes [9].

Here, we develop an experimental technique to measure the trap stiffness in plasmonic nano-optical tweezers based on the analysis of the fluorescence emission from the trapped nanoparticle. The main idea is that the fluorescence intensity fluctuations contain information about the movement of the trapped nanoparticle around its equilibrium position [30–32]. By analyzing these fluctuations and computing their temporal correlation, we can retrieve the trap stiffness (Fig. 1). This approach requires only a point detection of the total fluorescence intensity, hence an avalanche photodiode or a photomultiplier tube can be used instead of an EM-CDD or a sCMOS camera. As we will show further below, this approach based on fluorescence is equally sensitive independently of the spectral position of the plasmonic resonance. A limitation is of course that the trapped object must be fluorescent and of size smaller than the gap G, but for calibration and design optimization purposes there is a wide range of fluorescent nano-objects commercially available.

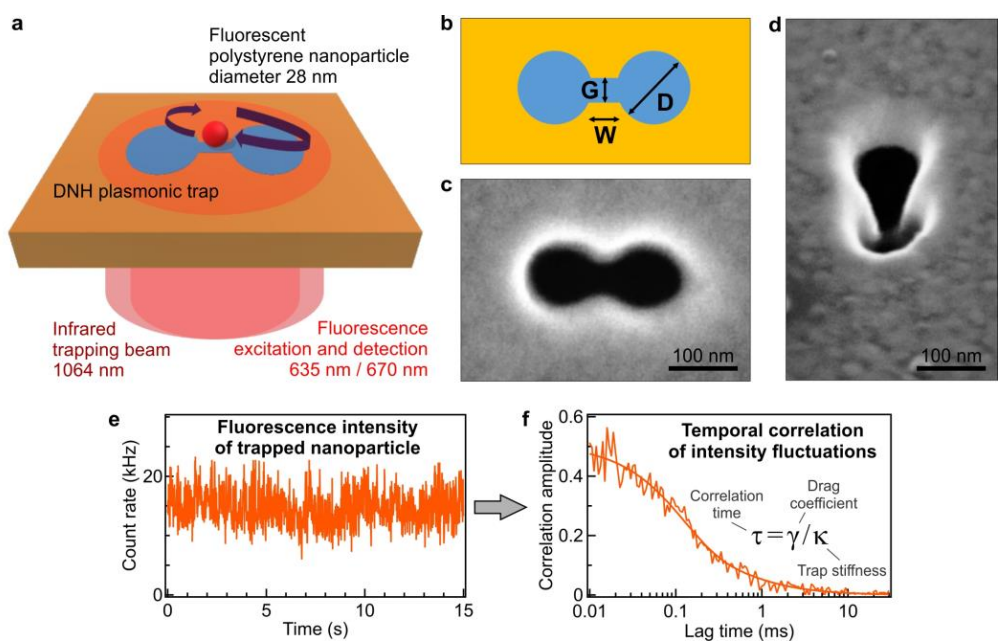

Fig. 1. (a) Trapping of a fluorescent polystyrene nanoparticle with a double nanohole aperture (DNH). The setup combines a 1064 nm laser beam for trapping with a confocal fluorescence microscope to excite and collect the fluorescence from the nanoparticle. (b) Design parameters used to describe the DNH aperture. The gold thickness is constant here at 100 nm. (c) Scanning electron microscope image of a DNH with D 80 nm, G 30 nm and W 30 nm. (d) Tilted view of the sample to better show the gap region narrowing down to 30 nm at the bottom of the DNH. (e)-(f) Concept to determine the plasmonic trap stiffness: the temporal fluctuations of the fluorescence intensity from the trapped nanoparticle (e) are analyzed by computing the temporal correlation (f). The characteristic correlation time $\tau$ corresponds to the Stokes drag coefficient $\gamma$ divided by the trap stiffness $\kappa$.

To demonstrate the effectiveness of this approach, we characterize plasmonic nano-tweezers based on double nanohole (DNH) apertures, which have received a large interest for plasmonic trapping [5,7,9,16,17,20,21,24,33]. We report the influence of the DNH geometry parameters directly on the experimental trap stiffness. Thanks to this characterization, we can optimize the DNH design and improve its performance by 10× over the previous state-of-the-art using a similar structure [7,9,22]. We also relate our observations to numerical simulations of the DNH local intensity enhancement and spectral response.

**Theory**

Here we describe the method to determine the trap stiffness based on the temporal correlation of the fluorescence intensity $I(t)$ recorded during a trapping event. The changes in the fluorescence intensity occur as the trapped nanoparticle explores different positions $r(t)$ inside the trap. We consider here that the trapped nanoparticle is a point-like source and that its intensity distribution follows a Gaussian profile

$$I(r(t)) = I_0 \prod_i e^{\frac{-r_i(t)^2}{2\omega_i^2}} \quad (1)$$

where $I_0$ is the peak intensity and $\omega_i$ is the width of the intensity distribution at 1/e² around the peak along the different directions $i = x, y, z$. For small displacements $r \ll \omega_i$, the intensity distribution is simplified [31]

$$I(r(t)) = I_0 (1 - \sum_i \frac{r_i(t)^2}{2\omega_i^2}) \quad (2)$$

We now introduce the intensity-normalized correlation function defined as $G(t) = \langle \delta I(0) \delta I(t) \rangle / \langle I(t) \rangle^2$, where $\delta I(t) = I(t) - \langle I \rangle$ is the intensity fluctuation around the average. Using Eq. (2), the intensity correlation depends on the nanoparticle position correlation $\langle r_i(0) r_i(t) \rangle$ [30,31]

$$G(t) = \frac{1}{2}\sum_i \frac{1}{\omega_i^4} \langle r_i(0)r_i(t)\rangle^2 \qquad (3)$$

The trapped nanoparticle follows a Brownian motion inside a harmonic potential $V = \frac{1}{2}\kappa_i r_i^2$ where $\kappa_i$ denotes trap stiffness along each direction. Its position obeys the Langevin equation:

$$\gamma \frac{dr_i}{dt} + \kappa_i r_i = f(t) \qquad (4)$$

where $\gamma$ is the Stokes drag coefficient and $f(t)$ represents an stochastic thermal force with $\langle f(t)\rangle = 0$ and $\langle f(t_1)f(t_2)\rangle = 2\gamma k_B T \delta(t_1 - t_2)$, where $k_B$ is Boltzmann's constant and $T$ the absolute temperature. The solution to the Langevin equation is given by [34]

$$r_i(t) = \frac{1}{\gamma}\int_{-\infty}^{t} dt' \exp\left(-\frac{t-t'}{\tau_i}\right) f(t') \qquad (5)$$

where $\tau_i = \gamma/\kappa_i$. The solution for $r_i(t)$ allows to compute the temporal correlation [30–32]

$$\langle r_i(0)r_i(t)\rangle = \frac{k_B T}{\kappa_i} \exp\left(-\frac{t}{\tau_i}\right) \qquad (6)$$

This equation can then be inserted into Eq. (3) to express the intensity correlation function:

$$G(t) = \frac{1}{2}\sum_i \frac{(k_B T)^2}{\kappa_i^2 \omega_i^4} \exp(-2\frac{t}{\tau_i}) \qquad (7)$$

Eq. (7) shows that the intensity correlation for a trapped nanoparticles decays as $\exp(-2t/\tau)$, where the correlation time $\tau = \gamma/\kappa$ amounts to the ratio of the Stokes drag coefficient $\gamma$ by the trap stiffness $\kappa$. Our approach to measure the trap stiffness is briefly summarized in Fig. 1e,f. The fluorescence intensity of the trapped nanoparticle is recorded to compute the correlation function $G$, fit the correlation time and then retrieve the trap stiffness as $\kappa_i = \gamma/\tau_i$. The last element needed is the Stokes drag coefficient $\gamma$ which is given by Faxen's law to account for the influence of the interface near the nanoparticle:[7,35]

$$\gamma = \frac{6\pi\eta R}{(1 - \frac{9}{16}\frac{R}{h} + \frac{1}{8}\left(\frac{R}{h}\right)^3 - \frac{45}{256}\left(\frac{R}{h}\right)^4 - \frac{1}{16}\left(\frac{R}{h}\right)^5)} \qquad (8)$$

where $h$ is the average distance between the center of the nanoparticle and the aperture wall, $R$ is the nanoparticle radius and $\eta$ the viscosity of the water medium. With $R = 14$ nm and $h = 15$ nm, the $6\pi\eta R$ term in the drag coefficient is increased by a factor 2.5, which appears similar to what was used in a previous study [7]. For a more accurate measurement, we also take into account the temperature dependence of the water viscosity $\eta$ using the Vogel equation [36] and our previous calibration of the temperature inside DNH structures as a function of the infrared laser power [37,38].

**Experimental results**

The trapped objects are dark red fluorescent carboxylate-modified polystyrene nanospheres (Thermofisher F8783) with a measured diameter of 28 nm. The fluorescent nanoparticles are diluted to a concentration of $4.25 \times 10^{11}$ particles/mL in ultrapure water with 2 mM Sodium dodecyl sulfate (SDS, Sigma Aldrich) used as the surfactant to avoid nanoparticles aggregation and ease trapping [33].

The double nanohole (DNH) design is summarized on Fig. 1b, and is defined by the aperture diameter D, the gap size G between the gold apex and the gap width W [24]. We directly mill the DNH apertures by gallium-based focused ion beam (FEI dual beam DB235 Strata) into a 100 nm thick gold layer with a 5 nm chromium adhesion layer deposited on a borosilicate glass coverslip [38]. The gallium beam current is set to 10 pA. Fig. 1c,d shows a typical sample with D 80 nm, G 30 nm and W 30 nm. The gap size G is measured at the bottom of the DNH at the glass-metal interface. The tapering of the gap is typical of FIB-based fabrication and its effect is taken into account in our numerical simulations.

We first explore the influence of the DNH diameter D. All the other DNH design parameters and experimental conditions are kept identical. Fig. 2a shows the fluorescence intensity correlation functions for different diameters and their fits using Eq. (9) described in the Method section. These traces are obtained while computing the temporal correlation of the fluorescence intensity recorded when the nanoparticle is trapped. The theory indicated by Eq. (7) predicts that both the correlation amplitude and correlation time decrease when the trap stiffness grows. The physical explanation behind this is that a larger trap stiffness implies movements of reduced amplitude (the trap is narrower), hence the nanoparticle goes back quickly to its equilibrium position. According to the theory, our approach could in principle distinguish the trap stiffness along the different spatial directions. However, the experimental correlation data becomes difficult to fit reliably for lag times shorter than 10 µs due to an increased statistical noise. Hence we focus here on the correlation in the sub-ms range.

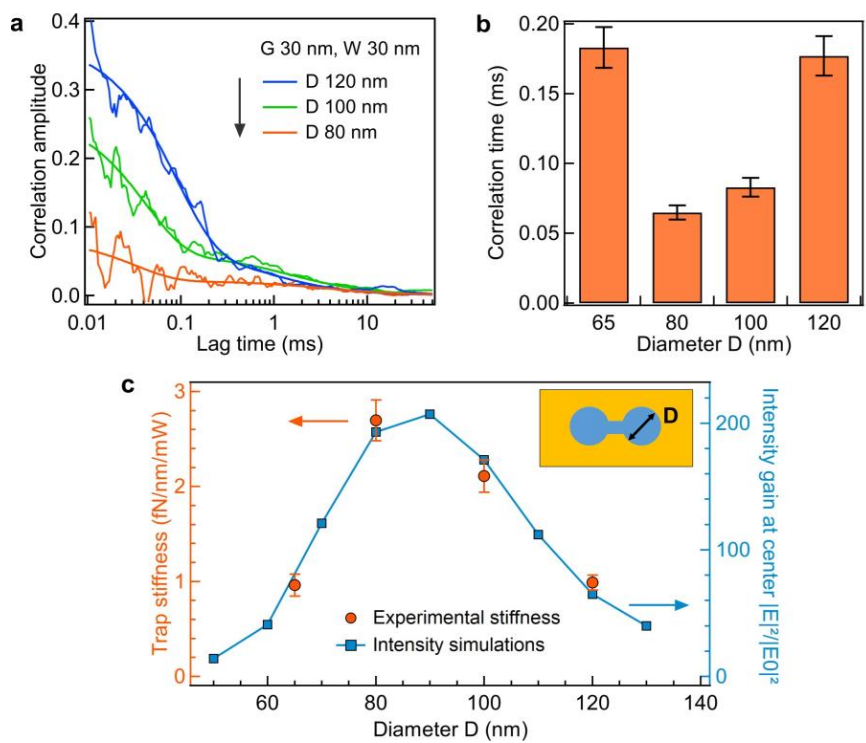

Fig. 2. Influence of the DNH diameter D. (a) Correlation functions (thin lines) and numerical fits (thick lines) for DNH of decreasing diameters D. The gap size G and width W are both 30 nm. The correlation function for D = 65 nm nearly overlaps with the one found for D = 120 nm and is not represented here for clarity. (b) Correlation times extracted from the fits in (a). (c) Evolution of the experimental trap stiffness (left axis, orange circles) as a function

of the DNH diameter. The same graph also represents the local intensity gain in the center of the DNH computed by FDTD at 1064 nm (right axis, blue squares).

From the correlation times in Fig. 2b and the computation of the drag coefficient with Eq. (8), we determine the trap stiffness. To compare it with other studies, we normalize the trap stiffness by the 3 mW laser power, so the stiffness is expressed in fN nm$^{-1}$ mW$^{-1}$. The values are summarized on Fig. 2c. Selecting the right aperture diameter can have a marked influence on the trap stiffness as we measure a nearly 3× larger stiffness for 80 nm diameter than for 65 or 120 nm. This clearly shows the importance of optimizing the DNH design to reach the best trapping performance.

Our experimental stiffness data follows also nicely the same trend as the numerical simulations of the infrared laser intensity in the center of the DNH (Fig. 2c). Here the electromagnetic field intensity is recorded at the center of the DNH gap (x = y = 0) and for the z position corresponding to the highest intensity along the vertical direction. While computing numerically the trap stiffness can be complicated, predicting the local electromagnetic intensity distribution at the laser wavelength is much easier as several numerical approaches and commercial softwares are readily available. This provides a simple approach to optimize the DNH design. Despite its simplicity, our experimental data show that computing the intensity enhancement at the center of the DNH is correct in optimizing the DNH design.

We next perform a similar study on the influence of the DNH gap width W. Fig. 3a shows the fluorescence correlation data from a trapped nanoparticle recorded for different W values. Here the diameter D is fixed to 80 nm according to the near optimum in Fig. 2c and the gap size G is kept at 30 nm to remain slightly larger than the trapped object. Again, we monitor a clear change in the correlation amplitude and characteristic time (Fig 3b), which indicate an optimum for the trap performance. The measured trap stiffness are plotted on Fig. 3c. They again follow a similar trend as the local 1064 nm intensity in the center of the DNH, which indirectly provides a confirmation for the validity of our measurements. Fig.

3c shows that the gap width also plays a critical role. We measure a 1.5× larger stiffness for W = 40 nm diameter than for 20 or 60 nm. Despite the diameter D is fixed to be near the optimum size according to Fig. 2c, choosing an improper width W can significantly and negatively affect the trap performance.

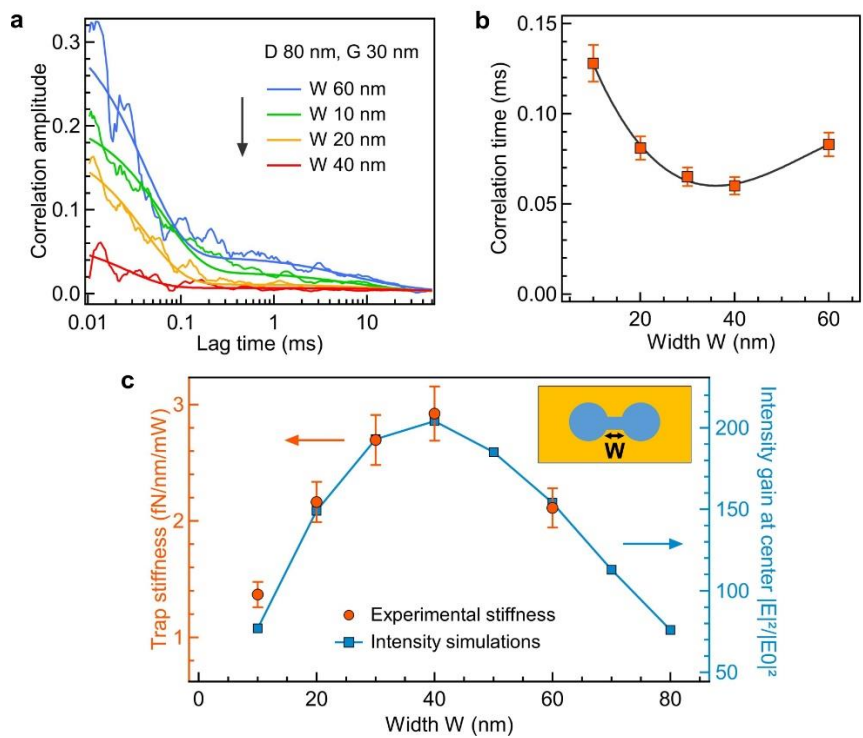

Fig. 3. Influence of the DNH gap width W. (a) Correlation functions (thin lines) and numerical fits (thick lines) for DNH of decreasing widths W. The diameter D is 80 nm and the gap size G is 30 nm for all these experiments. (b) Correlation times extracted from the fits in (a). the line is a guide to the eyes. (c) Evolution of the experimental trap stiffness (left axis, orange circles) as a function of the gap width. The local intensity gain in the center of the DNH computed by FDTD at 1064 nm is shown with blue squares (right axis).

## Discussion

While the gap G is set to be slightly larger than the size of the object of interest, the data in Fig. 2c and 3c show that both the DNH diameter and width are important parameters in maximizing the trap stiffness. The design D 80 nm, G 30 nm and W 40 nm provides the highest trap stiffness at 1064 nm wavelength. In Fig. 4, we relate this result to the spectral resonance of the DNH [20,21,24]. The design with D 80 nm, G 30 nm and W 40 nm corresponds to the case where the plasmonic resonance matches the 1064 nm laser wavelength (Fig. 4a,b). Larger diameters or widths lead to red-shifted resonances, while smaller dimensions yield blue-shifted resonances. For these non-optimum cases, the local laser intensity will be less enhanced, and the trap stiffness will not reach its highest value.

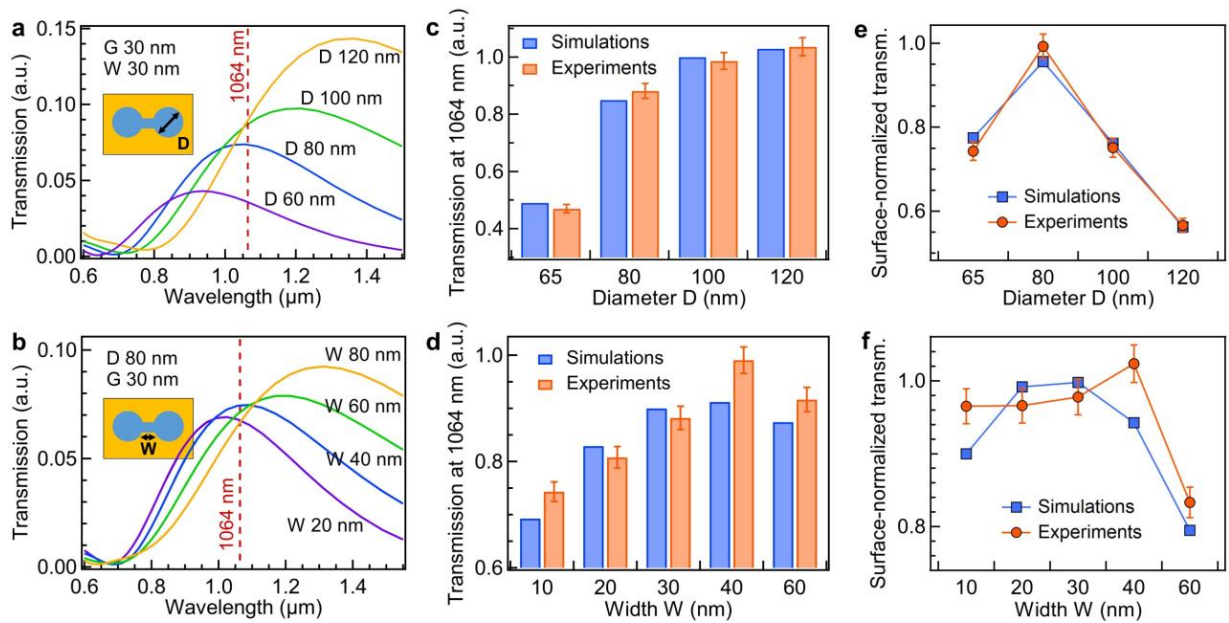

Fig. 4. Spectral influence of the DNH design parameters. (a) Numerical simulations of the DNH transmission spectrum for increasing diameters D. The gap size G and width W are both 30 nm. (b) Same as (a) for increasing gap widths W. The diameter D is 80 nm and the gap size G is 30 nm. (c,d) Comparison of the intensity transmission at 1064 nm between FDTD simulations and experiments for different diameters D (c) and widths W (d). (e,f) Transmission in (c,d) normalized by the DNH surface to better show the plasmonic resonance at 1064 nm.

Fig. 4a,b show an additional advantage of our characterization method. Measurements based on the infrared transmission usually select designs where the resonance is detuned from the trapping laser wavelength so as to yield a larger relative change in the transmission induced by the presence of the nanoparticle [5,22]. Our approach with the fluorescence emission separates the stiffness readout (here fluorescence in the red spectral region) from the trapping illumination (in the infrared). This works equally well when the DNH resonance matches the illumination wavelength to maximize the intensity enhancement and the trap stiffness.

By measuring the experimental transmission at 1064 nm and comparing the results to numerical simulations (Fig. 4c,d), we confirm the geometry and the size of each fabricated structures. The correct agreement between the experimental and simulated transmissions validate our fabrication process. Finally to better represent the occurrence of the plasmonic resonance, we normalize the transmission by the total geometrical surface of the DNH $(2\pi(D/2)^2 + W \cdot G)$. The case with D 80 nm, G 30 nm and W 40 nm corresponding to the plasmonic resonance matching the 1064 nm wavelength provides the highest surface-normalized transmission (Fig. 4e,f).

Regarding the trap stiffness and its comparison to the stat-of-the-art, we obtain an intensity-normalized trap stiffness of $2.9 \pm 0.2$ fN nm$^{-1}$ mW$^{-1}$ for the best DNH design with D 80 nm, G 30 nm and W 40 nm while trapping 28 nm polystyrene nanoparticles. To compare this value to other works using slightly different nanoparticle diameters, we further normalize our trap stiffness by a factor $\times (10 \, nm/R)^3$ so that the volume-scaled stiffness becomes $1.0 \pm 0.1$ fN nm$^{-1}$ mW$^{-1}$ in our case. This later value can now be compared with other studies where all stiffness are normalized to the case of a polystyrene nanoparticle with 20 nm diameter. While using a similar DNH geometry, Kotnala and Gordon reported a stiffness of 0.1 fN nm$^{-1}$ mW$^{-1}$ [7]. However, in their case, the DNH was not fully optimized with D 170 nm, G 25 nm and W 40 nm at a lower 820 nm laser wavelength than us. Crozier and coworkers also investigated a DNH geometry with D 100 nm, G 30 nm and W 40 nm at 1064 nm wavelength which appears even closer to our optimized DNH [9]. By monitoring the position of the fluorescent nanoparticle with an EM-CCD camera, they measured a stiffness of 0.04 fN nm$^{-1}$ mW$^{-1}$ but this measurement may potentially underestimate the real stiffness due to the limited frame rate of the EM-CCD. The numerical simulations in the same article [9] using Maxwell stress tensor indicate a maximum stiffness of 1.5 fN nm$^{-1}$ mW$^{-1}$ along the axis connecting each aperture center in the DNH and 3.7 fN nm$^{-1}$ mW$^{-1}$ along the direction parallel to the metal apex joining the apertures. These simulated values come closer to our direction-averaged experimental value. Looking to other plasmonic designs, Quidant and coworkers obtained a normalized trap stiffness of 0.12 fN nm$^{-1}$ mW$^{-1}$ with a resonant bowtie aperture [22] (the data was obtained while working with 60 nm gold nanoparticles, and the stiffness has been scaled according to Clausius-Mossotti relation to account for the change in the object polarizability). Saleh and Dionne found that coaxial apertures provide 0.36 fN nm$^{-1}$ mW$^{-1}$ [25]. Chormaic and coworkers obtained 0.25 fN nm$^{-1}$ mW$^{-1}$ using connected nanohole arrays [39] and 8.65 fN nm$^{-1}$ mW$^{-1}$ using asymmetric split-rings metamaterial [11] which is the highest value reported to date. While comparing the trap stiffness among different experiments, we should also mention that we use here 2 mM sodium dodecyl sulfate surfactant which improves the trap performance by ensuring that the thermophoretic force positively contributes to the net restoring force [33].

**Conclusions**

We have detailed an experimental method to measure the trap stiffness in a nano-optical tweezers based on the analysis of the temporal fluctuations of the fluorescence emitted by the trapped nano-object. This approach is not limited to the DNH geometry and can be applied to other plasmonic or all-dielectric structures designs. Among its different advantages, our approach can work equally well at the plasmonic resonance, does not require any expensive highly sensitive camera system, and remains computationally very simple. We demonstrate its effectiveness by characterizing the trap stiffness in different double nanohole apertures. The exploration of the design parameters allows us to optimize experimentally the trap stiffness, reaching a value that is 10× higher than the previous state-of-the-art using a similar structure [7,9,22]. Our experimental results are related to numerical simulations of the DNH local intensity enhancement and spectral response, providing design guidelines that will be useful to both theoreticians and experimentalists working on nano-optical tweezers. We also show that computing either the local intensity in the center of the trap or the surface-normalized transmission at the trap laser wavelength is sufficient to optimize the plasmonic design and get the highest trap stiffness.

**Methods**

**Optical microscope**

The optical microscope uses a continuous wave 1064 nm laser for trapping (Ventus 1064-2W). This infrared laser beam is overlapped with a 635 nm laser beam (Picoquant LDH-P-635) used to excite the fluorescence of the nanoparticles (Fig. 1a). The microscope objective is a Zeiss Plan-Neofluar with 40x magnification, 1.3 numerical aperture and oil immersion. The infrared laser power at the focus of the objective is 3 mW and the spot size is about 1 µm². This intensity enables correct trapping of single particles and remains low enough to avoid trapping several nanoparticles. The red laser power for fluorescence excitation is 5 µW. The fluorescence emission from the trapped nanoparticle is collected by the same objective and sent to a pair of avalanche photodiodes (Picoquant MPD-5CTC) after a 650-690 nm bandpass filter and a 30 µm pinhole conjugated to the microscope focus. The fluorescence intensity is recorded by a time correlated single photon counting module (Picoquant Picoharp 300 with time-tagged time-resolved option and PHR 800 router) allowing to compute the temporal correlation of the fluorescence time trace with the Symphotime 64 software (Picoquant).

**Correlation analysis**

The temporal correlation of the fluorescence intensity is fitted using the model:

$$G(t) = \rho_1 \exp\left(-2\frac{t}{\tau}\right) + \frac{\rho_2}{1 + \frac{t}{\tau_L}} \quad (9)$$

where $\rho_1$ and $\rho_2$ are the correlation amplitudes, $\tau = \gamma/\kappa$ is the correlation time used to determine the trap stiffness, and $\tau_L$ is an additional correlation time used to account for the long-term time fluctuations. While we focus on the first term in $\exp(-2t/\tau)$ to measure the trap stiffness, we find that a second term is needed to account for additional long-term fluctuations. The origin for this second term remains unclear and could be related to mechanical or thermal drifts as well as the influence of other nanoparticles in the solution interacting with the trapped nanoparticle [40].

**Numerical simulations**

The numerical simulations are performed using finite-difference time-domain (FDTD) with RSoft Fullwave software. The simulations are run with a 2 nm mesh size and a dwell time corresponding to 0.001λ. The convergence is checked after ten optical periods. The simulated geometry accounts for the conical shape of the apertures milled by focused ion beam [41]. The complex permittivity for gold is taken from ref [42].

**Funding sources**

This project has received funding from the European Research Council (ERC) under the European Union's Horizon 2020 research and innovation programme (grant agreement No 723241 TryptoBoost) and from the Agence Nationale de la Recherche (ANR) under grant agreement ANR-17-CE09-0026-01.

**Conflicts of interest**

The authors declare no competing financial interest.